\documentclass[aps,pre,twocolumn,amsmath,amssymb,superscriptaddress,longbibliography,showpacs,floatfix]{revtex4-2}
\usepackage{graphicx}
\begin{document}

\title{Critical exponents of correlated percolation of sites not visited by a random walk}
\author{Raz \surname{Halifa Levi}}\email{razhalifa@gmail.com}
\affiliation{The Faculty of Engineering,
Tel Aviv University, Tel Aviv 6997801, Israel}
\author{Yacov Kantor}
\affiliation{School of Physics and
Astronomy, Tel Aviv University, Tel Aviv 6997801, Israel}

\date{\today}

\begin{abstract}
We consider a $d$-dimensional correlated percolation problem of sites {\em not}
visited by a random walk on a hypercubic lattice $L^d$ for $d=3$, 4 and 5. The
length of the random walk is ${\cal N}=uL^d$. Close to the critical value $u=u_c$,
many geometrical properties of the problem can be described as powers (critical
exponents) of $u_c-u$, such as $\beta$, which controls the strength of the spanning
cluster, and $\gamma$, which characterizes the behavior of the mean finite cluster
size $S$. We show that at $u_c$ the ratio between the mean mass of the largest
cluster $M_1$ and the mass of the second largest cluster $M_2$ is independent of
$L$ and  can be used to find $u_c$. We calculate $\beta$ from the $L$ dependence
of $M_1$ and $M_2$, and $\gamma$ from the finite size scaling of $S$. The resulting
exponent $\beta$ remains close to 1 in all dimensions. The exponent $\gamma$
decreases from $\approx 3.9$ in $d=3$ to $\approx1.9$ in $d=4$ and
$\approx 1.3$ in $d=5$ towards $\gamma=1$ expected in $d=6$, which is
close to $\gamma=4/(d-2)$.
\end{abstract}

\maketitle

\section{\label{sec:intro}Introduction}

Percolation theory~\cite{Stauffer91,Gremmett99,Saberi15} provides a theoretical
framework for several classes of theories describing generation of long-range
connectivity from contacts between nearby objects. On regular lattices, these
objects are sites or bonds. A group of connected neighboring sites or bonds
forms a {\em cluster}. In this paper we consider {\em site} percolation on hypercubic
lattices. Such systems are characterized by a single parameter, such as the probability
$p$ of an occupied sites.  A cluster is {\em spanning} if it contains an uninterrupted
path between opposing boundaries in a specific direction. In finite systems of linear
size $L$ the probability $\Pi(p,L)$ of the existence of such a path gradually increases
with increasing $p$. For an infinite system $\Pi(p,\infty)$ becomes a step-function
$\Theta(p-p_c)$ jumping from 0 to 1 at a particular {\em critical concentration} $p_c$.
The mean spatial extent (linear size) of {\em finite} clusters is called the correlation
length $\xi$.  Close to $p_c$ it diverges as
\begin{equation}\label{eq:nu}
\xi\sim|p-p_c|^{-\nu}.
\end{equation}

One of the simplest percolation models is Bernoulli site percolation on a
$d$-dimensional lattice, where each lattice site is {\em independently} occupied
with probability $p$. The exponent $\nu=\nu_{\rm B}$ of this problem decreases
with increasing $d$ reaching $\nu_{\rm B}=\frac{1}{2}$ for $d\ge d_c=6$, i.e.,
at and above the {\em upper critical dimension} $d_c$~\cite{Toulouse74}. Besides
$\nu$, the percolation problem is characterized by a
large number of additional {\em critical exponents}, such as $\gamma$
describing the divergence of the mean finite cluster size $S$ (an accurate definition will
be provided in the Sec.~\ref{sec:clusters}) in the vicinity of $p_c$:
\begin{equation}\label{eq:gamma}
S\sim |p-p_c|^{-\gamma},
\end{equation}
or the fraction of sites $P$ belonging to the infinite cluster for $p>p_c$, also called the {\em strength} of the infinite cluster:
\begin{equation}\label{eq:beta}
P\sim (p-p_c)^{\beta}.
\end{equation}

For Bernoulli percolation, the values of these and other exponents are known exactly
for $d=2$ and $d\ge6$, and have accurate numerical estimates for $d=3$, 4, and 5.
There are numerous equalities relating various exponents describing the behavior close to
the threshold, such as the hyperscaling relation
\begin{equation}\label{eq:hyper}
d\nu=2\beta+\gamma
\end{equation}
and others~\cite{Stauffer91}. As a result, the values of many exponents can be deduced
from only two exponents. Nevertheless, due to the limited accuracy of the numerical
studies, additional exponents are measured and the values of the results are verified
using the known relations.

In Sec.~\ref{sec:model} we define a correlated percolation model and briefly
overview its properties as well as some known results. In Sec.~\ref{sec:clusters}
we define the main quantities and briefly describe the numerical methods used in this
paper. Section~\ref{sec:M1M2} describes the properties of the largest clusters
and their use to verify the position of the percolation threshold. Critical
exponent $\beta$ characterizing the strength of the infinite cluster is found from
the $L$ dependence of the first and the second largest clusters in Sec.~\ref{sec:beta}, while
the exponent controlling the behavior of the mean cluster size $\gamma$ is found in
Sec.~\ref{sec:gamma}. Summary Sec.~\ref{sec:summary} briefly discusses the
results.

\section{Long-range model}\label{sec:model}

Minor modifications of the Bernoulli percolation model, such as the introduction of
short-range correlations between sites, consideration of bond percolation,
or even power-law correlations $\sim 1/r^b$ between present sites at distance $r$
with {\em large} power $b$  do not change the critical behavior of the
system ~\cite{Weinrib84}. However, if $b<2/\nu_{\rm B}$, then the correlations
are relevant, and $\nu_{\rm B}$ is replaced~\cite{Weinrib84} by
\begin{equation}\label{eq:long_range}
\nu_{\rm long\ range}=2/b\ .
\end{equation}

There is a variety of studies of correlated percolation
models~\cite{Coniglio09,Gori17,Riordan2011,DSouza15,K_PRB33,Schrenk13,Schrenk16}.
In this paper, we consider a problem, where an initially
full $d$-dimensional hypercubic lattice of linear size $L$ (in lattice constants)
and volume $L^d$ has its sites removed by an ${\cal N}$-step random walk
(RW) on the lattice. The {\em length} of the RW that starts at a random position is
proportional to the {\em volume} of the lattice, namely, ${\cal N}=uL^d$. Periodic
boundary conditions are imposed on the walk on a finite lattice, i.e., the walker
exiting through one boundary of the lattice reemerges on the opposite
boundary. The parameter $u$ controls the length of the RW  and the fraction $p$ of
unvisited sites. In repeated realizations of this process at fixed $u$, the average
$p$ is a monotonically decreasing function of $u$. For large $L$ (and $d\ge 3$) there
is a simple relation between these quantities:  $p=\exp(-A_d u)$, where $A_d$ are
known constants (see Ref.~\cite{KK_PRE100} and references therein). An experimental
realization of this model involves a gel of crosslinked polymers with the random walker
represented by an enzyme that breaks the crosslinks between polymers that it
encounters~\cite{Berry00,Fadda03} eventually breaking the spanning cluster and
turning the solid gel into a liquid. The object of our study is the
sites {\em not visited} by the RW, that represent the surviving crosslinks.

\begin{table}[t]
  \begin{center}
    \begin{tabular}{|c|c|c|c|c|r|}
     \hline \hline
      $d$  & $\nu_{\rm th}$ & $\nu_{\rm num}$ & $u_c$ &  $L_{\rm max}$\\
     \hline \hline
     3 &      2         & $2.04\pm0.08$  & $3.15\pm0.01$  & 512\\ \hline
     4 &      1         & $1.0\pm 0.1$   & $2.99\pm0.01$  &  64\\ \hline
     5 & $\frac{2}{3}$  & $0.65\pm0.03$  & $3.025\pm0.008$&  32\\ \hline
     6 & $\frac{1}{2}$  & $\sim 0.6$     & $3.10\pm0.05$  &  16\\ \hline
     \hline
    \end{tabular}
  \end{center}
\caption{Some results of previous studies: First two columns show the space dimension
$d$, and the value of $\nu$ predicted by Weinrib~\cite{Weinrib84}, as applied to our problem~\cite{KK_PRE100} in Eq.~\eqref{eq:nu_RW}, respectively.
The 3rd and the 4th columns provide numerical estimates of $\nu$ and
$u_c$ from \cite{KK_PRE100}, while the last column gives the maximal linear size
of a lattice used in that study.}
  \label{tab:known_values}
\end{table}

The variable $u$ naturally replaces $p$ in this problem, and $p_c$ is replaced by the
critical value $u_c$. (Keep in mind that the system percolates, i.e., has spanning
clusters of unvisited sites {\em below} $u_c$.) In a previous study, it has been found
that on hypercubic lattice for $3\le d\le 6$ the threshold values
$u_c\approx 3$~\cite{KK_PRE100}. (See Table~\ref{tab:known_values}.) This problem has
been previously studied by
Banavar {\em et al.} in $d=2$ and $3$~\cite{Banavar85}, while Abete {\em et al.}
considered the critical behavior near the percolation threshold in $d=3$~\cite{Abete04}.
More recently, Kantor and Kardar studied the percolation properties of the
problem for $2\le d\le 6$~\cite{KK_PRE100}, and showed that the problem
has no percolation threshold in $d=2$ (for more details, see Ref.~\cite{FK_PRE103}).
Recently, we received a work by Chalhoub {\em et al.} \cite{Chalhoub24} with a detailed
numerical study of the critical properties of this problem in large systems in $d=3$ and
analytical predictions regarding this problem and several similar problems in general $d$.
We will compare their results with ours wherever it is appropriate.

\begin{figure}[t]
\includegraphics[width=8 truecm]{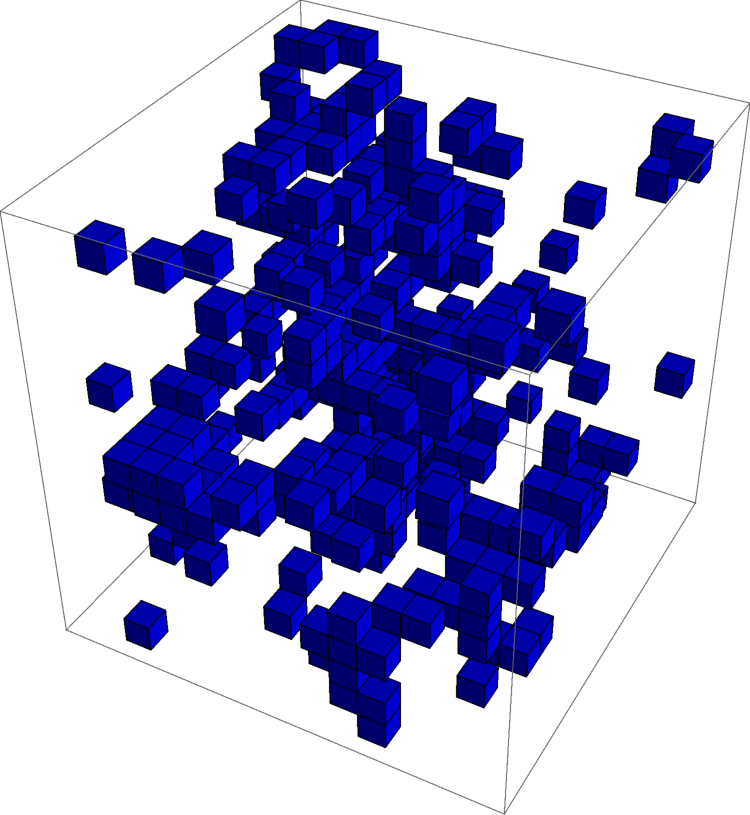}
\caption{Three-dimensional percolation in a small system of linear size $L=16$. The sample was
created by a random walk of ${\cal N}=uL^3$ steps, where $u=4$. The blue cubes show the
{\it unvisited} sites, which form several clusters. The visited sites form a single cluster.}
\label{fig:3D_percolation_u4}
\end{figure}

Figure \ref{fig:3D_percolation_u4} depicts a typical configuration of the
percolation of sites (blue cubes) not visited by a RW  in $d=3$. In this
realization, the number of steps of the RW exceeds four times the
number of available lattice positions. Nevertheless, a significant fraction
of the sites remains unvisited, since the RW frequently revisits previously visited
sites. All {\em visited} sites form a {\em single cluster} since they have been
created by a single RW. The unvisited sites form many clusters of various sizes.
The picture is quite different from the usual geometries of Bernoulli percolation.
The RW has a fractal dimension $\tilde{d}_f=2$.  On an {\em infinite} lattice an
${\cal N}$-step walk explores distance
$r\sim {\cal N}^{1/\tilde{d}_f}={\cal N}^{1/2}$. Within that distance, the density of
sites visited by the RW is ${\cal N}/r^d\sim 1/r^{d-2}$. On a {\em finite}
lattice, a RW that exits through one boundary and re-enters through the
opposite boundary creates almost uncorrelated strands, and the sites on
different strands are no longer power-law correlated. A RW of length $uL^d$
creates approximately $uL^{d-2}$ such strands, that contribute to an uncorrelated
background density of sites. However, at distances $r$ smaller than the
lattice size, there remains a residual correlation of sites that happen to be on the
{\em same} strand of the RW. Consequently, the {\em cumulant} of the correlation
(from which the overall background density has been subtracted) still has
$\sim 1/r^{d-2}$ correlation like a RW on an infinite lattice.
Consider a random variable $v(\vec{x})$ which is 1 if the site at position $\vec{x}$
is {\em unvisited} by the RW, and zero otherwise. It is complementary to the
variable representing the {\em visited} site (their sum is 1) and therefore it
has the same cumulant: $\langle v(\vec{x})v(\vec{y})\rangle_c\sim 1/|\vec{x}-\vec{y}|^{d-2}$~\cite{KK_PRE100}. Thus, the correlation power is
$b=d-2$, and the relation \eqref{eq:long_range} becomes
\begin{equation}\label{eq:nu_RW}
\nu=2/(d-2),\ \ {\rm for}\ \ 3\le d\le 6.
\end{equation}
We note that recently Feshanjerdi {\em et al.} \cite{Feshanjerdi23} studied a
three-dimensional ``carbon copy" version of our problem, where they considered
percolation of sites {\em visited} by a (modified) RW. Their model is expected to have similar
critical properties.

Kantor and Kardar's study of the problem in $3\le d\le6$~\cite{KK_PRE100},
focussed on the behavior of the percolation probability $\Pi(u,L)$ as a
function of RW parameter $u$ and system size $L$. As $L$ increases,
$\Pi(u,L)$ approaches the step function $\Theta(u_c-u)$. Thus, by
examining the $u$ dependence of $\Pi$ for a sequence of $L$ values,
it was possible to estimate the transition point $u_c$, while by
studying $L$ dependence of the width of the transition between percolating
($\Pi\approx 1$) and nonpercolating ($\Pi\approx 0$) states it was
possible to determine the exponent $\nu$.
Table \ref{tab:known_values} compares numerically calculated values of
$\nu$~\cite{KK_PRE100} with the predictions of Weinrib~\cite{Weinrib84}
in Eq. \eqref{eq:nu_RW} and provides the calculated values of the thresholds.
(In $d=3$  an earlier estimate of $\nu$ on smaller systems ($L_{\rm max}=60$) was $1.8\pm0.1$~\cite{Abete04}, i.e., it slightly deviated from the predicted value of 2.)
The conclusion of Ref.~\cite{KK_PRE100} was that the numerical results validated the
theoretical prediction of  Weinrib~\cite{Weinrib84} in $d=3$, 4, and 5, and we
will use the theoretical values of $\nu$ in the current paper.
(Dimension $d=6$ is expected to be the upper critical dimension where $\nu=\frac{1}{2}$
coincides with $\nu_{\rm B}$ of Bernoulli percolation. The system sizes
in Ref.~\cite{KK_PRE100} were too small to reliably determine the value of the exponent.)

\section{Clusters and finite size scaling}\label{sec:clusters}

In a system of $N$ sites with $N_s$ clusters containing $s$
sites, we define {\em cluster density} $n_s\equiv N_s/N$. The total number
of occupied sites on a lattice $M=\sum_s sN_s$, and therefore the
fraction of occupied sites is $p=M/N=\sum_s sn_s$. It is natural
to define $n_s$ and $p$ as averages over an ensemble of systems, such as
$n_s\equiv \langle N_s\rangle/N$, where $\langle\cdot\rangle$ defines
defines ensemble average of ``$\cdot$". For simplicity
of the notation, we will omit the $\langle\rangle$ signs where their
presence is self-evident from the context. We also consider the thermodynamic
limit of infinite lattices $N\to\infty$, where all $n_s$ and $p$ approach
a finite limit. In an infinite lattice above the percolation threshold
$p_c$, an infinite cluster is present and occupies some fraction $P>0$ of
the sites. Since summation over all integer numbers $s$ does not include
an infinite cluster, the $\sum_s sn_s=p-P$. (Below the threshold, $P=0$ and
the formula reduces to the expression that we had before.) In finite
systems, the distinction between the ``would-be-infinite" cluster and
other clusters blurs and the transition itself is smeared. Therefore, it
is customary~\cite{Stauffer91} in numerical estimates of $N_s$ on finite
lattices to exclude the largest cluster of mass $s_{\rm max}\equiv M_1$
of any random realization of the system: For very large systems above
$p_c$ this is equivalent to the exclusion of the
infinite cluster, while below $p_c$ it slightly decreases the finite cluster
size, but this effect decays with increasing $L$. Values of $M_1$ are used
to estimate the fraction of the ``infinite" cluster $P=M_1/N$. This
procedure produces a nonvanishing effective $P$ below the threshold, but
its value decays to zero as the system size increases.

In infinite systems, the {\em second moment} $\chi\equiv\sum_ss^2n_s$
characterizes the size (mass) of the {\em finite} cluster, since the sum
excludes the infinite cluster, if such is present. In fact the mean size
$S$ of a cluster to which a given occupied site belongs, is given by
$S=\chi/(p-P)$~\cite{Stauffer91}. (In an infinite system both the mean
cluster size $S$ and the second moment $\chi$ diverge near $p_c$ as shown
in Eq.~\eqref{eq:gamma}, and both can be used for numerical estimates of
$\gamma$.) The expression for $\chi$ can be re-written as
$\chi=\sum_s s^2N_s/N=\sum_\alpha s_\alpha^2/N$, where the last sum
simply represents a sum of squares of sizes of each distinct cluster
$\alpha$ in the system. (In the finite systems in each configuration
in our calculations, we exclude the largest cluster from that sum.)

For the problem of the sites unvisited by a RW, the system percolates
{\em below} $u_c$ and in all the expressions describing the behavior of
$S$ and $P$ the arguments $p-p_c$ should be replaced by $u_c-u$, if
the sign of the expression matters.

The second moment $\chi$ or the mean cluster size $S$ are the geometrical analogs
of the susceptibility in magnetic systems. In an infinite system near
the threshold Eq.~\eqref{eq:gamma} can be rewritten as $S\sim |u-u_c|^{-\gamma}$,
and, similarly, Eq.~\eqref{eq:beta} becomes $P\sim (u_c-u)^{\beta}$.
The correlation length in Eq.~\eqref{eq:nu} can be rewritten
as $\xi\sim|u-u_c|^{-\nu}$.  In a system of linear size $L$, the
correlation length is naturally truncated by the system size and, consequently,
$S$ stops increasing when the expression for diverging $\xi$  exceeds $L$.
Similarly, $P$, which was supposed to vanish at $u_c$ abandons its power-law
decay when $\xi\sim L$. In general, a critical quantity $X$ that was supposed
to be singular as $|u-u_c|^{-x}$, with either positive or negative exponent $x$
becomes finite when $\xi>L$. It is also possible that the apparent position of
the singularity (such as the position of the peak of $S$) may slightly differ from
$u_c$ and can be treated as an effective percolation threshold $u_c^*(L)$.
(Its actual value may depend on the quantity which is being considered.)
$u_c^*(L)$ keeps moving towards $u_c$ with
increasing $L$. It is expected that
\begin{equation}\label{eq:ucL}
|u_c-u_c^*(L)|\sim L^{-1/\nu}\ .
\end{equation}
Finite-size scaling theory was originally developed for thermodynamic
systems~\cite{Privman84} and later adapted to percolation
systems~\cite{Stauffer91}. Since the behavior of the system is controlled
by $\xi/L\sim |u-u_c|^{-\nu}/L$, it is convenient to describe the system
by using parameter $v=(u-u_c)L^{1/\nu}$. In terms of $v$ the position of
the peak of $S$ becomes independent of $L$. When $v$  is small, i.e.,
$|v|<V$, where $V$ is some model-dependent number of order unity, the
behavior of the system is controlled by the finite size $L$, while for
$|v|\gg V$ we expect to recover infinite-system behavior. In general,
the behavior of a critical quantity $X$ can be described by
\begin{equation}\label{eq:gX}
X=L^{x/\nu}g_X[L^{1/\nu}(u-u_c)]=L^{x/\nu}g_X(v)\ ,
\end{equation}
where $g_X(v)$ is a scaling function. Thus, by drawing $XL^{-x/\nu}$
{\em vs.}~$v$ with properly chosen parameters $u_c$, $\nu$ and $x$ it should
be possible to collapse all the results into a single curve, which is the
function $g_X(v)$, as long as both $L$ and $\xi$ are significantly larger
than the lattice constant. To recover the $L=\infty$ behavior
$X\sim |u-u_c|^{-x}$, we must have $g_X(v)\sim |v|^{-x}$ for $|v|\gg V$.
For small $v$ the scaling function approaches some constant
$g_X(0)$ leading to  $u$-independent result $X\sim L^{x/\nu}$.

The numerical procedure used in this paper is rather straightforward:
Initially, all sites on a hypercubic lattice $L^d$ are present. The system
sizes in $d=3$ are $L=16$, 32, 64,\dots, 512. For $d=4$, they are 16, 32, 64
and 128, while in $d=5$ the $L$s are increased by approximately factor
$\sqrt{2}$: 16, 23, 32, 45. These $L$s are slightly larger than in the
previous study (see Table~\ref{tab:known_values}). A RW starts at a randomly
selected site and for
a given $u$ performs $uL^d$ steps removing the sites that have been visited.
At each realization, the Hoshen-Kopelman algorithm \cite{Hoshen76} is
used to identify {\em all} clusters, and from the complete list of clusters
all necessary quantities characterizing that configuration can be calculated.
For a fixed $u$ and $L$ this RW procedure is repeated 1000 times and various
quantities are averaged over the realizations. For each $L$, a
sequence of $u$s is studied to obtain $u$ dependence of such averages at typical
steps $\Delta u=0.05$, except for specific situations mentioned in the text.

\section{Two largest clusters}\label{sec:M1M2}

\begin{figure}[t]
\includegraphics[width=8.5 truecm, clip=true]{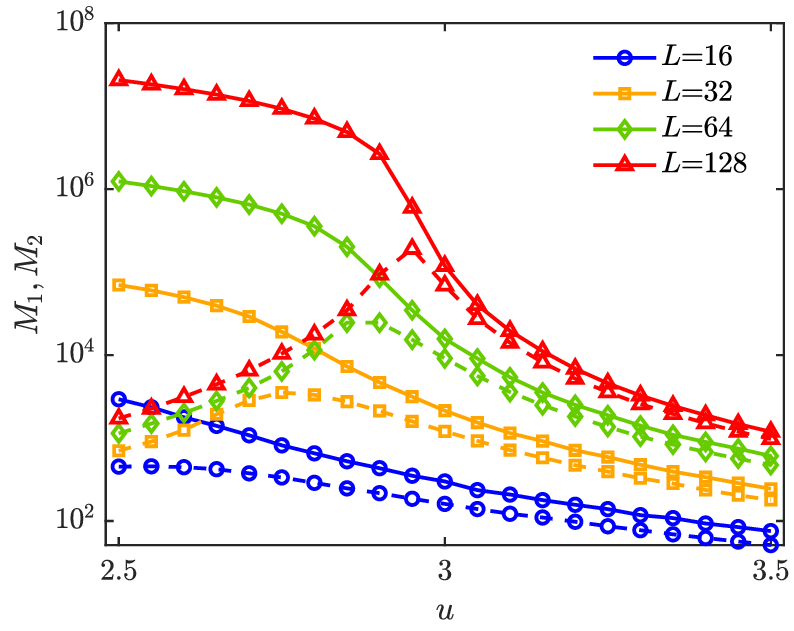}
\caption{Semilogarithmic plot of the mean mass
$M_1$ of the largest cluster (solid lines) and the mean mass $M_2$ of the
second largest cluster (dashed lines) as a function of RW parameter $u$, in
$d=4$ for (bottom to top) $L=16$ (blue circles), 32 (yellow squares), 64
(green diamonds) and 128 (red triangles). Here, as in other graphs, each point
represents an average over 1000 configurations. Relative statistical
errors in  $M_1$ and $M_2$ are $\lesssim 4\%$, which on logarithmic
scale results in error bars $\Delta \log_{10}M_2\lesssim 0.02$ which are
smaller than the symbol sizes.
Horizontal separation between data points $\Delta u=0.05$.
Qualitatively similar behavior was observed in $d=3$ and $d=5$.
}
\label{fig:M1andM2}
\end{figure}

At any $d$, on a finite lattice of linear size $L$ the typical mass
(number of sites) $M_1$ of the {\em largest} cluster has a very different
$u$ dependence than the mass $M_2$ of the {\em second largest} cluster:
In the percolating region ($u<u_c$), $M_1$ is essentially the mass of
the ``would-be-infinite" cluster $PL^d$. For a very small $u$, we have
$M_1\approx L^d$ and it decreases with increasing $u$. For $u>u_c$ in the
nonpercolating region, $M_1$ is just the largest of the finite clusters
that are present and it keeps decreasing with increasing $u$. So, overall,
$M_1$ is a monotonically decreasing function of $u$. For a very small
$u$ almost all space is occupied by the massive spanning cluster,
and $M_2\approx 1$. As $u$ increases (still in the percolating region), so
do the sizes of finite clusters and $M_2$ increases. Above $u_c$ the
typical cluster sizes decrease with increasing $u$ and so does the
mass $M_2$. In that region, it has a similar behavior to $M_1$, but,
obviously, is smaller than $M_1$. We expect $M_2$ to have a maximum
somewhere around $u_c$. Figure \ref{fig:M1andM2} shows $M_1$ and $M_2$
for several system sizes $L$ in $d=4$. (Similar behavior is found at
other $d$s.) Indeed, $M_1$ is a monotonically decreasing function, while
$M_2$ has a maximum at some $u_c^*(L)$, which approaches $u_c$ with
increasing $L$ as in  Eq.~\eqref{eq:ucL}.

At the percolation threshold, the largest cluster has fractal structure,
and its mass increases with the system size as $M_1\sim L^{d_f}$
\cite{Stauffer91}, where $d_f$ is the fractal dimension of
the infinite cluster. However, it has been known for quite some time
\cite{Margolina82,Jan98,Jan99} that for Bernoulli percolation at the threshold,
the mass of the {\em second} largest cluster $M_2$ (as well as third and
higher rank clusters) also scales with the same power, although it has a
smaller prefactor.  Therefore, one can expect that the {\em ratios} of two
such cluster masses, represented either by the ensemble average
$\langle M_1/M_2\rangle$, or by $\langle M_1\rangle/\langle M_2\rangle$
will scale with the zeroth power of $L$.
So this ratio can be treated using the
same finite size arguments that were used for $X$ and lead to scaling form
\eqref{eq:gX}, with $x=0$ and a different scaling function~\cite{Silva00}
$g_0$. While this reasoning originally applied to Bernoulli percolation,
it originates from the argument that at a critical point in the absence of
length scale both the largest and the second largest clusters share the
same behavior. Therefore, this property can be expected (subject to
verification) in our correlated percolation problem. In such a case, we expect
\begin{equation}\label{eq:g0}
\langle M_1/M_2\rangle\ {\rm or}\ \langle M_1\rangle/\langle M_2\rangle
=g_0[L^{1/\nu}(u-u_c)]=g_0(v)\ .
\end{equation}

At $u=u_c$ this equation means that the ratio of masses of the first and
the second largest clusters at the critical point should be $g_0(0)$, i.e.,
independent of $L$. The presence of such an intersection at $u_c$ both
confirms that the first and the second largest clusters have the same
fractal dimension, and enables an alternative approach for an accurate
determination of the percolation threshold. This method has been
successfully used by da Silva {\em et al.}~\cite{Silva02} to locate
$p_c$ in Bernoulli percolation problem.

\begin{figure}[t]
\includegraphics[width=8.5 truecm, clip=true]{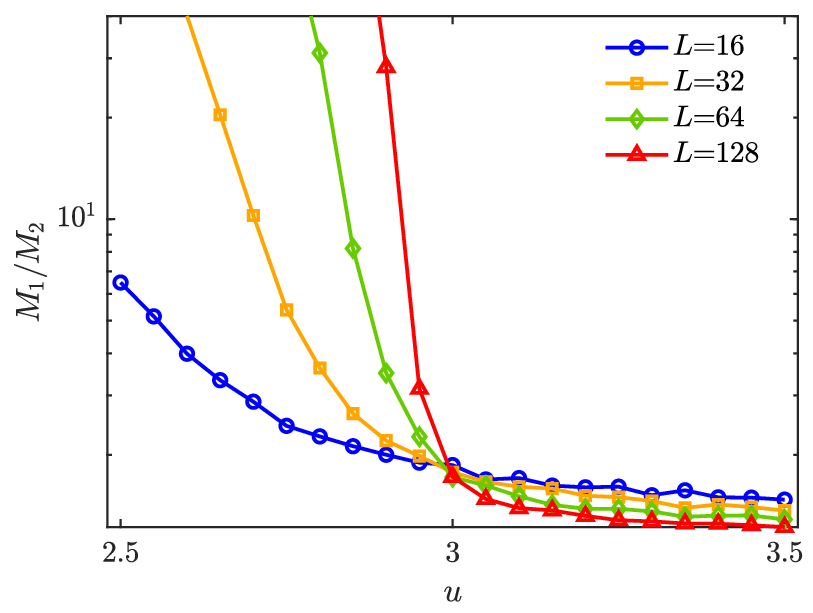}
\caption{Semilogarithmic plot of the ratios of mean mass $M_1$ of the largest
cluster and the mean mass $M_2$ of the second largest cluster as a function
of RW length parameter $u$ in $d=4$ for system sizes (left to right on the top parts
of the graphs) $L=16$
(blue circles), 32 (yellow squares), 64 (green
diamonds), and 128 (red triangles). This figure uses the same data as
Fig.~\ref{fig:M1andM2}, and shares the same technical details. The intersection
points of subsequent-$L$ lines are between 2.99 and 3.00 and provide an estimate
for $u_c$ (Smaller $\Delta u$s
were used close to $u_c$ but the extra data points are not shown here.)
Qualitatively similar behavior was observed in $d=3$ and $d=5$.
}
\label{fig:M1ratioM2}
\end{figure}

The study in Ref.~\cite{KK_PRE100} of correlated percolation extracted
the value of $u_c$ by examining the percolation probability $\Pi(u,L)$ as
a function of $u$ for several values of $L$, and observing the approach
of that function to the infinite-$L$ limit of step function $\Theta(u_c-u)$.
With increasing $u$ the function $\Pi(u,L)$ monotonically decreases
from a percolating small-$u$ regime to a nonpercolating large-$u$ region.
The effective transition point $\tilde{u}$ was defined as a point where
$\Pi(\tilde{u},L)=c$, with an arbitrary constant $0<c<1$. Clearly,
the value of $\tilde{u}$ depends both on $L$ and $c$. Since in the
$L\to\infty$ limit the $\Pi$ becomes a step function,
in that limit $\tilde{u}\to u_c$ independently of the choice of $c$. Indeed,
in Ref.~\cite{KK_PRE100} it was shown that various choices of $c$ lead
to similar $L\to\infty$ extrapolations. However, all $\tilde{u}$ values exhibited
significant $L$ dependence, and therefore it is beneficial to use an alternative
method to confirm that threshold values appearing in Table~\ref{tab:known_values}.
We apply the method of da Silva {\em et al.}~\cite{Silva02} to our problem.
Figure~\ref{fig:M1ratioM2} depicts the $u$ dependence of $M_1/M_2$ for various
values of $L$ in $d=4$. We note that the intersection points between various
lines are concentrated between 2.99 and 3.00, even for moderate $L$s, similar
to $u_c=2.99\pm0.01$ that has been obtained by strong extrapolations in
Ref.~\cite{KK_PRE100}. (We will use value of $u_c=2.995$ for the fits in $d=4$ in the remainder
of the paper.)

We performed similar studies of $M_1/M_2$ in $d=3$ and $d=5$ and confirmed
the known values of $u_c$ listed in Table~\ref{tab:known_values}. (Fig. 1 in
Ref.~\cite{Chalhoub24} uses an analogous method of line-intersection for different
types of quantities to determine $u_c$ in $d=3$ and concludes that our
original value $u_c=3.15$ was correct.) We will use the $u_c$ values
from Table~\ref{tab:known_values} (for $d=3$ and 5) in the remainder of this article.
While the results obtained in the following sections require the knowledge
of $u_c$, they are rather insensitive to its exact value and even changes in
the values of $u_c$ as large as 0.01 or 0.02 do not modify the estimates of
the critical exponents.

\section{Exponent $\beta$}\label{sec:beta}

Critical exponent $\beta$ is defined by the behavior of the infinite cluster $P$
near the percolation threshold, i.e., in an infinite system $P\sim (u_c-u)^\beta$,
for $u<u_c$. One can directly plot measured  $P$ {\em vs.} $u_c-u$ on a logarithmic plot for
a sequence of increasing $L$s, and determine the exponent $\beta$. For a finite
$L$ the power law is truncated when $\xi$ reaches $L$, and it is possible to treat that
truncation more systematically by considering a finite size scaling form
for $P$ which is analogous to Eq.~\eqref{eq:gX}
\begin{equation}\label{eq:gP}
P=L^{-\beta/\nu}g_P[L^{1/\nu}(u-u_c)]=L^{-\beta/\nu}g_P(v)\ ,
\end{equation}
where $g_P(v)$ is a scaling function, which for large {\em negative} $v$
should have the behavior $g_P(v)\sim (-v)^{\beta}$ to recover
the desired $L$-independent behavior of $P$, while for
small $v$ we obtain $u$-independent result $P\sim L^{-\beta/\nu}$. Since
$P$ is small close to $u_c$ it increases the relative statistical errors
and reduces the accuracy of the numerical calculations near that point.
On the other hand, one can concentrate on a fixed $v$, such as $v=0$, and
explore the relation $P=L^{-\beta/\nu}g_P(0)$. Since $P=M_1/L^d$, we may
use the relation $M_1\sim L^{d_f}$, with the fractal dimension
\begin{equation}\label{eq:df}
d_f=d-\beta/\nu\ .
\end{equation}

\begin{figure}[t]
\includegraphics[width=8.5 truecm, clip=true]{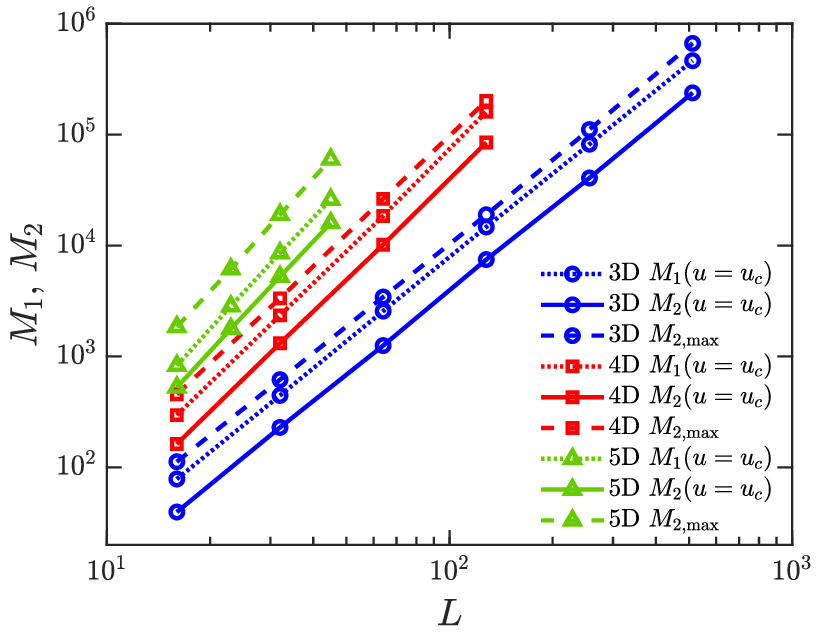}
\caption{Logarithmic plots of  the mean mass $M_1$ of the largest cluster
measured at $u_c$ ($v=0$) (dotted lines), and of the mean mass $M_2$ of
the {\em second} largest cluster measured at $u_c$ ($v=0$) (solid lines)
or at the maximum of the curve ($v=v_{\rm max}$) (dashed lines) as a function
of $L$ for (bottom to top triplets of lines) $d=3$ (blue circles), $d=4$
(red squares), and $d=5$ (green triangles). Each data point depicts an ensemble
average and the error bars, as explained in Fig.~\ref{fig:M1andM2}, are smaller
than the symbol sizes. Slopes of the curves are the fractal dimensions $d_f$ of
the clusters.
}
\label{fig:M2vsL}
\end{figure}

Since both $M_1$ and $M_2$ have the same fractal dimension
at $u_c$, by plotting  both $M_1(v=0)$ and $M_2(v=0)$ as a functions of $L$
on a logarithmic scale, one can determine $d_f$, and, consequently, the
exponent $\beta$. The solid lines in Fig.~\ref{fig:M2vsL} depict on a
logarithmic scale the dependence of $M_2$ at $u_c$ on system size $L$, in
the complete range of $L$s used for each $d=3$, 4, and 5. From the slopes of the lines ($d_f$)
and from Eq.~\eqref{eq:df} we deduce $\beta=0.98\pm0.04$ in $d=3$,
$\beta=1.00\pm0.06$ in $d=4$, and $\beta=1.09\pm0.05$ in $d=5$.
Similarly, the dotted lines depict the values of $M_1$, from
which we deduce $\beta=0.99\pm0.02$ in $d=3$,
$\beta=0.98\pm0.10$ in $d=4$, and $\beta=1.11\pm0.07$ in $d=5$. The
stated errors combine small statistical errors of individual $M_1$ or $M_2$ data
points (relative errors smaller than 4\%) with the scatter of the points
around the straight line. We could not determine a systematic $L$ dependence
of the slope, which might indicate that we are measuring an effective
exponent which changes with increasing $L$. If such a change was present,
it was masked by the statistical fluctuations of the slope.

\begin{table}[b!]
  \begin{center}
    \begin{tabular}{|c|c|c|r|l|}
     \hline \hline
      $\nu_{\rm num}$ & $\beta$ & $\gamma$ &  $L_{\rm max}$ &Ref\\
     \hline \hline
     $1.8\pm0.1$  &   $1.0\pm0.1$  & $3.5\pm0.2$  &   60& \cite{Abete04} \\ \hline
     $1.99\pm0.01^*$  &$0.99\pm0.01^*$&           & 1024& \cite{Feshanjerdi23}\\ \hline
     $1.95\pm0.11$   &             & $4.00\pm0.36$      & 600&\cite{Chalhoub24}\\
     $2.02\pm0.08$  &   $0.98\pm0.08$  &        &    & \\ \hline
     \hline
    \end{tabular}
  \end{center}
\caption{Numerical values of exponents from previous studies of percolation
of sites or bonds {\em not visited} by a RW in $d=3$. The fifth column provides the source reference,
while the fourth column gives the maximal linear size of a lattice used in that study.
(This table excludes the results appearing on the first line of Table \ref{tab:known_values}.) Asterisks indicate results for a (presumably equivalent) model of sites {\em visited} by a modified
RW. The values of $\beta$ or $\gamma$ are frequently calculated as the ratios $\beta/\nu_{\rm num}$
or $\gamma/\nu_{\rm num}$, respectively, and have been calculated from the results appearing
in the references. Some authors used several methods to calculate the same exponents, and the reader
should consult the original references.
}
  \label{tab:num_exponents}
\end{table}

The result for $d=3$ should be compared with the result $\beta=1.0\pm0.1$ of
Abete {\em et al.}~\cite{Abete04} obtained on smaller systems ($L_{\rm max}=60$). (In that
calculation, a slightly smaller value of $\nu$ was used.) Our result in $d=3$
is also consistent (within quoted errors) with the result obtained in
Ref.~\cite{Chalhoub24} for larger systems ($L_{\rm max}=600$). A similar result
was also obtained by Feshanjerdi {\em et al.}~\cite{Feshanjerdi23} for a problem
of {\em visited} sites; they conjectured that the exact value of $\beta$ in
$d=3$ should be the integer value 1.
Table \ref{tab:num_exponents} displays previously calculated
values of $\beta$ and other exponents in $d=3$ by several authors.
Note, that our $\beta$ values in $d=3$ are consistent
with the previously known values.

As an additional check of the presence or absence of systematic errors,
we measured $M_2$ at its maximal value. We verified that, within statistical
uncertainty, the maximal values of $M_2$ for all $L$ appear at the same
value of $v_{\rm max}$. This is
correct for each $d$, although the actual definition of $v$ depends on
$d$ since it involves specific $u_c$ and $\nu$. Thus, the measurement of
$M_2(v=v_{\rm max})$ for various $L$s should produce the same $L$ dependence
$M_2\sim L^{d_f}$, although with a larger prefactor, as we got for
$v=0$ (or $u=u_c$). This method of measurement introduces an additional
source of errors, since finding maximal $M_2$ requires interpolation
between data points close to that maximum. The presence
of this additional error increase in the estimated statistical
errors. Dashed lines in Fig.~\ref{fig:M2vsL} depict $M_2$ as a function
of $L$ measured at the points of maxima. The measured slopes ($d_f$)
produce slightly different values: $\beta=0.99\pm0.07$ in $d=3$,
$\beta=1.07\pm0.06$ in $d=4$, and $\beta=1.13\pm0.05$ in $d=5$.
While the results at $v=v_{\rm max}$ are less reliable,
they are similar to the previous sets of $\beta$s.

\begin{figure}[t]
\includegraphics[width=8.5 truecm, clip=true]{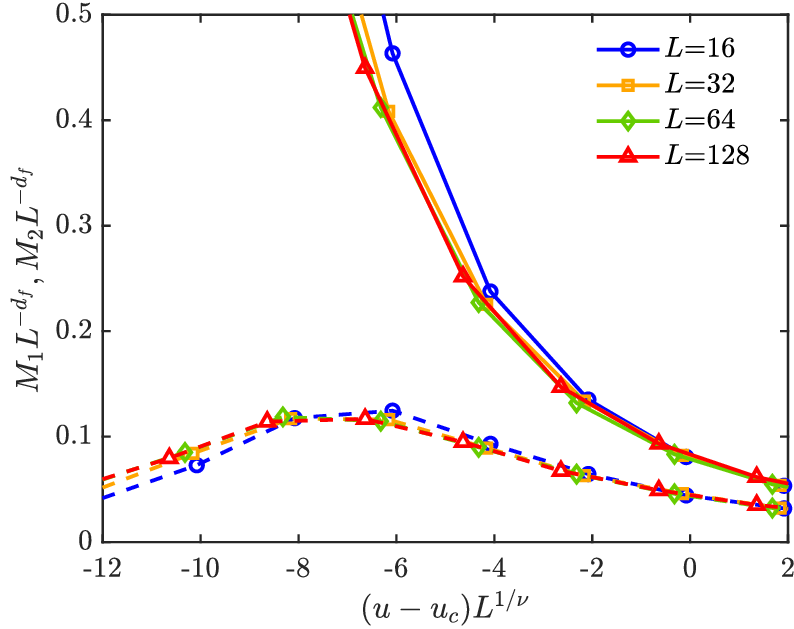}
\caption{Data-collapse plot of the scaled cluster sizes $M_1L^{-d_f}$
(top solid lines) and $M_2L^{-d_f}$ (bottom dashed lines)
in $d=4$ {\em vs.}~scaled parameter $v=(u-u_c)L^{1/\nu}$. This graph was obtained for
$d_f$ with $\beta=1.04$ [see Eq.~\eqref{eq:df}], and it remains a good fit when
$\beta$ is changed by $\pm 0.04$. This graph is a scaled form of
Fig.~\ref{fig:M1andM2}. The error bars are smaller than the sizes of the symbols.
}
\label{fig:Mvsv4Dscaled}
\end{figure}

The study of the behavior of $M_1$ and $M_2$, demonstrated in this Section,
supports the extension of the scaling form in Eq.~\eqref{eq:gP} to $M_i$ ($i=1,2$),
namely
\begin{equation}\label{eq:gM}
M_i=L^{d_f}g_{M_i}[L^{1/\nu}(u-u_c)]=L^{d_f}g_{M_i}(v)\ ,
\end{equation}
where $g_{M_1}(v)\equiv g_P(v)$. Figure~\ref{fig:Mvsv4Dscaled} demonstrates
such scaling via data collapse of scaled $M_1$ and $M_2$ for various $L$s for $d=4$.
[In this graph, as well as in other scaled plots (the last two figures of the paper)
instead of sampling the functions at fixed intervals $\Delta u$, we use (almost) fixed
intervals $\Delta v$ of the scaled variable.)
We note, that the value of $\beta$ used in this fit is in the middle of $\beta$
values obtained from Fig.~\ref{fig:M2vsL} for the same $d$. Analogous behavior
is observed in other dimensions.

As mentioned at the beginning of this section, the behavior of
$g_P(v)$ for {\em large} negative $v$ is consistent with the expected
power law but is not accurate enough to determine the value of $\beta$.

We note, that in $d=6$, the exponent of our problem should coincide with
the exponent of Bernoulli percolation and have mean field value $\beta=1$.
The fact that $\beta$ maintains the value so close to 1 in all dimensions
makes one wonder if $\beta=1$ is the exact dimension-independent value of
$\beta$. Such conjecture has been made by Feshanjerdi
{\em et al.}~\cite{Feshanjerdi23} in $d=3$ (for a system of visited sites)
and theoretically suggested to be valid in other dimensions by Chalhoub
{\em et al.}~\cite{Chalhoub24} (see Table I in their work).

\begin{figure}[t]
\includegraphics[width=8.5 truecm, clip=true]{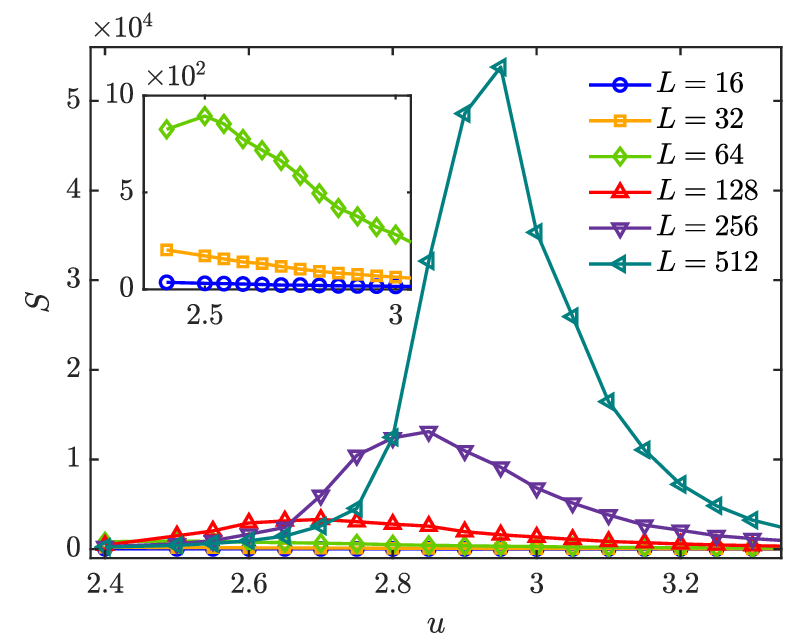}
\caption{Mean cluster size $S$
{\em vs.}~$u$ in $d=3$ for $L=16$, 32,\dots,512. Error bars are
smaller than the symbols. The inset shows the graphs for the three smallest $L$s
 with an expanded vertical scale. The expected critical point is
$u_c=3.15$, while the maxima of the curves appear at slightly lower
values $u^*_c(L)$ that drift towards $u_c$ with increasing $L$.
}
\label{fig:Svsu3D}
\end{figure}

\section{Exponent $\gamma$}\label{sec:gamma}

\begin{figure}[t]
\includegraphics[width=8.5 truecm, clip=true]{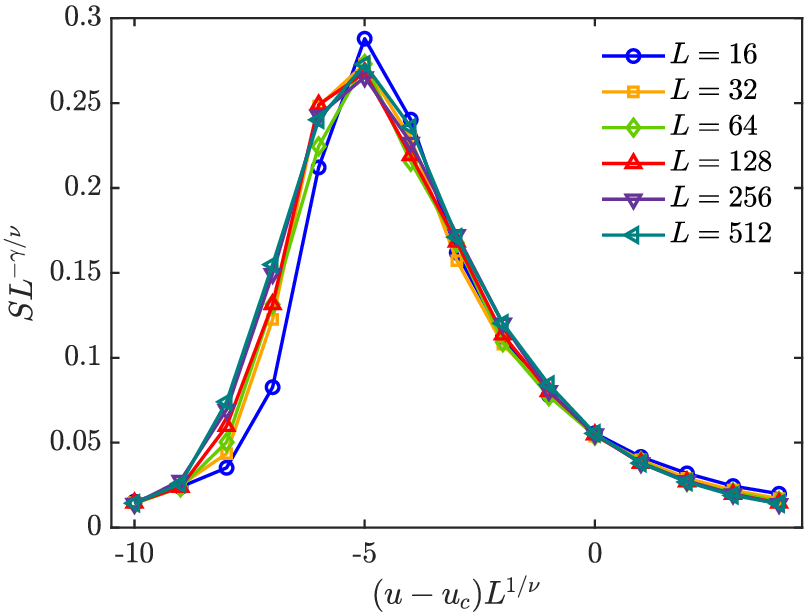}
\caption{Data-collapse plot of the scaled mean cluster size $SL^{-y}$
in $d=3$ {\em vs.}~scaled parameter $v=(u-u_c)L^{1/\nu}$.
This graph was obtained for $\gamma=y\nu=3.90$ and it remains a good fit when
$\gamma$ is changed by $\pm 0.05$. This graph is a scaled form of
Fig.~\ref{fig:Svsu3D}.
The error bars are smaller than the sizes of the symbols.
}
\label{fig:Svsu3Dscaled}
\end{figure}

Mean cluster size $S$ is one of the most important characteristics of
percolation. In infinite systems, similarly to the susceptibility
in magnetic systems, it diverges at the critical point as in
Eq.~\eqref{eq:gamma}, and it would be possible to measure the exponent
$\gamma$ by directly plotting $S$ as a function of the distance from the
critical point on a logarithmic scale. (Exponent $\gamma$ is expected to be
the same both above and below that point.) However, in a finite system, the
divergence is truncated when $\xi$ exceeds $L$.
Figure~\ref{fig:Svsu3D} depicts $S$ measured in $d=3$ as a function of $u$
for system sizes $L$ ranging from 16 to 512. As $L$ increases, the height of
the peak in $S$ increases approximately as $L^2$. The position of the peak
appears at some $u_c^*(L)$,  which is smaller than $u_c$ and this effective
critical point shifts towards $u_c$ with increasing $L$ as indicated by
Eq.~\eqref{eq:ucL}, while the height of the peak increases almost as $L^2$.
As with other critical quantities, we will use scaled variable $v=(u-u_c)L^{1/\nu}$
and describe the behavior of $S$ using scaling form analogous to
Eq.~\eqref{eq:gX}, namely
\begin{equation}\label{eq:gS}
S=L^{\gamma/\nu}g_S[L^{1/\nu}(u-u_c)]=L^{\gamma/\nu}g_S(v)\ ,
\end{equation}
where $g_S(v)$ is a scaling function. To recover the $L=\infty$ behavior
$S\sim |u-u_c|^{-\gamma}$, we must have $g_S(v)\sim |v|^{-\gamma}$ for
$|v|\gg V$. For small $v$ the scaling function approaches some constant
$g_S(0)$, leading to  $u$-independent result $S\sim L^{\gamma/\nu}$.
Instead of focusing on $v=0$, we can look at a point $v=v_{\rm max}$, where
all $S$ reach their maxima independently of $L$. At this point, we also
expect $S\sim L^{\gamma/\nu}$ but with a larger prefactor. In fact, the
exponent $\gamma$ can be determined from the entire function  function
$g_S(v)$: We need to plot $S L^{-\gamma/\nu}$ {\em vs.} $v$. Such a
plot should produce  the function $g_S(v)$. The graphs for several
$L$s should collapse into a single plot, provided the fitting parameter
$y\equiv\gamma/\nu$ has been properly selected. In general, we should
use three fit parameters $\gamma$, $\nu$ and $u_c$. However, the latter
two are well known to us, see Table~\ref{tab:known_values}. (The fitting
is insensitive to small changes in $u_c$.) We therefore, use the known values of
$\nu=\nu_{\rm th}=2$ and $u_c=3.15$ from Table~\ref{tab:known_values}
and adjust only the parameter $y$ that depends on $\gamma$.
Using the above values, Fig.~\ref{fig:Svsu3Dscaled} presents the same results as Fig.~\ref{fig:Svsu3D} in a scaled form where the horizontal axis uses
variable $v=(u-u_c)L^{1/\nu}$. The vertical axis represents the scaled cluster
size $SL^{-y}$, with the fit parameter $y$ corresponding to
$\gamma=3.90$. The fit remains good up to a shift of $\pm 0.05$ in $\gamma$.
Note, that the best-fit value of $y$ is very close to 2. The obtained value of
$\gamma$ should be compared with $\gamma\approx 3.5$ by Abete
{\em et al.}~\cite{Abete04} obtained on smaller systems ($L_{\rm max}=60$)
and with a slightly smaller numerical value of $\nu$, and $\gamma\approx 4.0$
obtained by Chalhoub {\em et al.}~\cite{Chalhoub24} on a slightly larger system.
(See previous results in Table \ref{tab:num_exponents}.)

Our result was extracted from the behavior for small $v$ behavior
of $g_S(v)$. As expected, for large $|v|$ we observe a power law
which is consistent with $|v|^{-\gamma}$, but the quality and the range of the
data is too small for an accurate determination of the power.

\begin{figure}[t]
\includegraphics[width=8.4 truecm, clip=true]{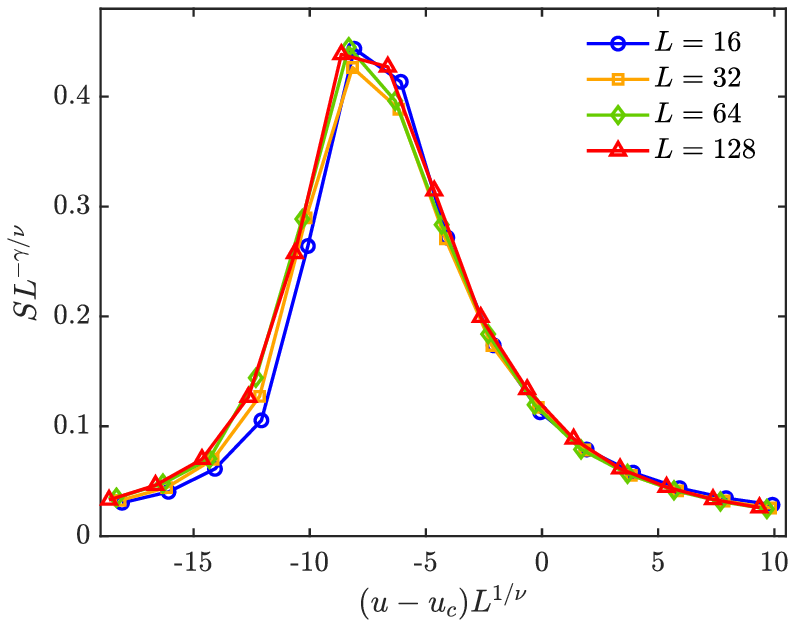}
\caption{Data-collapse plot of the scaled mean
cluster size $SL^{-y}$ in $d=4$ {\em vs.}~scaled parameter $v=(u-u_c)L^{1/\nu}$.
It is analogous to $d=3$ graph in Fig.~\ref{fig:Svsu3Dscaled}.
The exponent $y$ was adjusted to get the best data collapse. This
graph was obtained for $\gamma=y\nu=1.90$ and it remains a good fit when
$\gamma$ is changed by $\pm 0.05$.
The error bars are smaller than the symbol sizes.
}
\label{fig:Svsu4Dscaled}
\end{figure}

We repeated the above calculations in $d=4$ and $d=5$.
The decreasing range of $L$ leads to decreasing reliability of the fit.
As an example, we present Fig.~\ref{fig:Svsu4Dscaled} that shows
the scaled results in $d=4$, for $L$ ranging from 16 to 128. By fitting $y$
we find $\gamma=1.90\pm0.05$, and again parameter $y$ is very close to 2.
Similar analysis was performed in $d=5$ on systems of even smaller
linear size $L$ ranging from 16 to 45. The collapse plot (not shown) has slightly
lower quality and the best fit leads to the estimate $\gamma=1.30 \pm 0.05$.

As in the three-dimensional case we compared the values of $\gamma$ obtained
for small values of $v$ with possible power law behavior $|v|^{-\gamma}$.
Again, we obtained a reasonable correspondence for large negative values of
$v$, although the accuracy of those results was much lower. (We did not see
a clear power law for large positive $v$.)

We expect the exponents $\beta$ obtained in the previous section and the
exponents $\gamma$ obtained in this section to satisfy the hyperscaling
relation \eqref{eq:hyper}. Indeed, by inserting the calculated values the
relation is approximately satisfied if we allow for errors in the exponents.
In fact, we can use \eqref{eq:hyper} in an opposite way: if we substitute $\beta=1$
into the equation, and use $\nu$ from \eqref{eq:nu_RW} we find that in our problem
\begin{equation}\label{eq:gamma_th}
\gamma=4/(d-2)\ .
\end{equation}
We can see that this expression is very close to the values of $\gamma$
obtained in all dimensions. [We also note that \eqref{eq:gamma_th} leads for all
$3\le d\le 6$ to the exponent $y=\gamma/\nu=2$.]

\section{Discussion}\label{sec:summary}

The problem of sites not visited by a RW presents a relatively
simple case of correlated percolation. In this paper we calculated
exponents $\beta$ and $\gamma$ in $d=3$, 4, and 5, with reasonable
accuracy, using system sizes comparable with the previous works.
(There were no previous results in $d=4$ and 5.) We used
different methods to measure the exponents: measurement of the fractal
dimension of the first and the second largest clusters for calculation
of $\beta$ and finite size scaling for $\gamma$. Our results hint at simple
numerical values for the critical exponents, and one may hope to derive these
values from simple geometrical considerations.

\end{document}